\documentclass[twocolumn]{aastex631}

\usepackage{enumitem}
\usepackage{mathtools}

\def\MJ{M_\mathrm{J}}
\def\Mp{M_\mathrm{p}}

\def\RJ{R_\mathrm{J}}
\def\Rp{R_\mathrm{p}}
\def\Teff{T_{\text{eff}}}

\def\Rcav{R_\mathrm{cav}}
\def\Rout{R_\mathrm{out}}

\DeclareRobustCommand{\rchi}{{\mathpalette\irchi\relax}}
\newcommand{\irchi}[2]{\raisebox{\depth}{$#1\chi$}} 

\def\chimrs{\rchi^2_\mathrm{MRS}}

\begin{document}

\title{Mid-Infrared Spectrum of the Disk around the Forming Companion GQ Lup B Revealed by JWST/MIRI}

\author[0000-0001-7255-3251]{Gabriele Cugno}
\affiliation{Department of Astronomy, University of Michigan, Ann Arbor, MI 48109, USA}

\author[0000-0001-8718-3732]{Polychronis Patapis}
\affiliation{Institute for Particle Physics and Astrophysics, ETH Zurich,
Wolfgang-Pauli-Str. 27, 8093, Zurich , Switzerland}

\author[0000-0003-4335-0900]{Andrea Banzatti}
\affiliation{Department of Physics, Texas State University, 749 N Comanche Street, San Marcos, TX 78666, USA}

\author{Michael Meyer}
\affiliation{Department of Astronomy, University of Michigan, Ann Arbor, MI 48109, USA}

\author[0000-0002-5476-2663]{Felix A. Dannert}
\affiliation{Institute for Particle Physics and Astrophysics, ETH Zurich \\
Wolfgang-Pauli-Str. 27, 8093, Zurich , Switzerland}
\affiliation{National Center of Competence in Research PlanetS, Gesellschaftsstrasse 6, 3012 Bern, Switzerland}


\author[0000-0002-5823-3072]{Tomas Stolker}
\affiliation{Leiden Observatory, Leiden University, Niels Bohrweg 2, 2333 CA Leiden, The Netherlands}

\author[0000-0003-4816-3469]{Ryan J. MacDonald}
\affiliation{Department of Astronomy, University of Michigan, Ann Arbor, MI 48109, USA}

\author[0000-0001-7552-1562]{Klaus M. Pontoppidan}
\affiliation{Jet Propulsion Laboratory, California Institute of Technology, 4800 Oak Grove Drive, Pasadena, CA 91109, USA}



\begin{abstract}

GQ~Lup~B is a forming brown dwarf companion ($M\sim10-30~\MJ$) showing evidence for an infrared excess associated with a disk surronding the companion itself. 
Here we present mid-infrared (MIR) observations of GQ~Lup~B with the Medium Resolution Spectrograph (MRS) on JWST, spanning 4.8--11.7 $\mu$m. We remove the stellar contamination using reference differential imaging based on principal component analysis (PCA), demonstrating that the MRS can perform high-contrast science.
Our observations provide a sensitive probe of the disk surrounding GQ~Lup~B.
We find no sign of a silicate feature, similar to other disk surrounding very low mass objects, which likely implies significant grain growth ($a_{\mathrm{min}}\gtrsim5~\mu$m), and potentially dust settling. Additionally, we find that if the emission is dominated by an inner wall, the disk around the companion might have an inner cavity larger than the one set by sublimation. Conversely, if our data probe the emission from a thin flat disk, we find the disk to be very compact. More observations are required to confirm this findings and assess the vertical structure of the disk.
This approach paves the path to the future study of circumplanetary disks and their physical properties. Our results demonstrate that MIR spectroscopic observations can reveal the physical characteristics of disks around forming companions, providing unique insights into the formation of giant planets, brown dwarfs and their satellites. 

\end{abstract}



\section{Introduction} \label{sec:intro}

The direct detection of protoplanets still embedded in their circumstellar disk material can reveal unique insights into the formation processes sculpting exoplanetary systems. In particular, high-contrast imaging can constrain the brightness and location of planets forming in the disk, which can be related to disk substructures observed in the radio wavelengths and in optical and near-infrared (NIR) scattered light \citep{Andrews2020, Avenhaus2018, Ren2023}. Combining protoplanets and disk observations allows to study planet-disk interactions \citep[e.g.,][]{Bae2023} and to understand their complex interplay. At the interface between planets and circumstellar disks, circumplanetary disks (CPDs, \citealt{WardCanup2010}, referred here as the disk surrounding a companion, being it a planet or a brown dwarf) play a pivotal role in shaping the planetary environment by facilitating mass accretion, angular momentum transfer, and potential disk clearing.

To date, the only confirmed protoplanets orbit the young star PDS70 and are both undergoing accretion, carving a large cavity in the circumstellar disk \citep{Keppler2018, Haffert2019}. Since their discovery, the two protoplanets have been subject to studies using many instruments and observational techniques \citep[e.g.,][]{Muller2018, Wang2021, Cugno2021} to reveal their physical and chemical properties. Given their accreting nature \citep{Wagner2018, Haffert2019, Zhou2021}, the community has initiated an effort to detect their CPD.
Observations with the Atacama Large Millimeter/submillimiter Array (ALMA) revealed first a tentative and subsequently a distinct detection around PDS70c \citep{Isella2019, Benisty2021}. While these ALMA observations offer clear evidence for a CPD around PDS70~c, only a tentative indication of a CPD has been seen around PDS70~b in the form of excess emission in $M'$ band photometry \citep{Stolker2020_pds70, Christiaens2024}. 

The presence of IR excess emission at $\lambda\gtrsim5~\mu$m has been observed in several young planetary and brown dwarf mass companions \citep{Wu2017b, Martinez2022, Stolker2020_pds70, Stolker2021}. This type of emission has been associated with circumplanetary disk material and was often complemented with the detection of accretion tracers \citep[e.g.,][]{Wolff2017, SantamariaMiranda2018}, millimeter emission \citep[e.g.,][]{Wu2020, Wu2022} and polarimetric signals \citep{vanHolstein2021}. Despite these detections, little is known about the physical properties of circumplanetary disks, how they evolve and how they affect the formation of bound companions.

Among the young companions presenting indications of a circumplanetary disk is GQ~Lup~B. GQ~Lup~B, first discovered by \cite{Neuhauser2005}, orbits at a separation of $0\farcs7$ from the young ($2-5$~Myr, \citealt{MacGregor2017}) T~Tauri star GQ~Lup~A located at $d=154\pm0.7$~pc \citep{Gaia2022}. Several works tried to infer the companion mass, with values ranging from 10 to 40~$\MJ$ \citep[e.g.,][]{Marois2007, Seifahrt2007}, and its orbital parameters \citep[e.g.,][]{Ginski2014, Schwarz2016}. In particular, \cite{Stolker2021} fitted 15 yr of astrometric data and confirmed the large misalignment ($84\pm9^\circ$) between the orbit of GQ~Lup~B and the circumstellar disk suggested by \cite{Wu2017b}. They also constrained the semimajor axis of the orbit ($a=117^{+24}_{-23}$~au) and its eccentricity ($e=0.24^{+0.32}_{-0.17}$). Moreover, they modeled GQ~Lup~B's atmosphere, using optical and NIR spectra, finding it consistent with a low gravity object with $T_{\mathrm{eff}} = 2700$~K and a large planet radius of $\sim$3~$\RJ$. They identified an excess emission in VLT/NaCo NB4.05 and $M'$ data, which they tentatively attributed to a circumplanetary disk. Moreover, GQ~Lup~B shows signs of active accretion, with several hydrogen recombination lines detected in the optical and NIR \citep{Seifahrt2007, Zhou2014, Wu2017, Stolker2021, Demars2023}, pointing to a dynamic companion formation environment. Despite the detection of MIR excess and emission lines associated with accretion, ALMA observations at radio wavelength did not detect the circumplanetary disk \citep{Wu2017, MacGregor2017}.

In this work, we present the first spectrum of a circumplanetary disk surrounding the low-mass brown dwarf companion GQ~Lup~B, taken with the Mid InfraRed Instrument \citep[MIRI,][]{Wright2023} onboard the James Webb Space Telescope \citep[JWST,][]{Gardner2023}. We provide constraints on the properties of its disk, proposing a path to characterize CPDs of young protoplanets in the future. This Letter is structured as follows: in Sect.~\ref{sec:obs_red} we present our observations and data reduction, in Sect.~\ref{sec:results} we report the extracted spectrum of the companion and fit it with simple models. We discuss the results in Section~\ref{sec:discussion} and present our conclusions in Sect.~\ref{sec:conclusion}.

\begin{figure*}
    \centering
    \includegraphics[width = \textwidth]{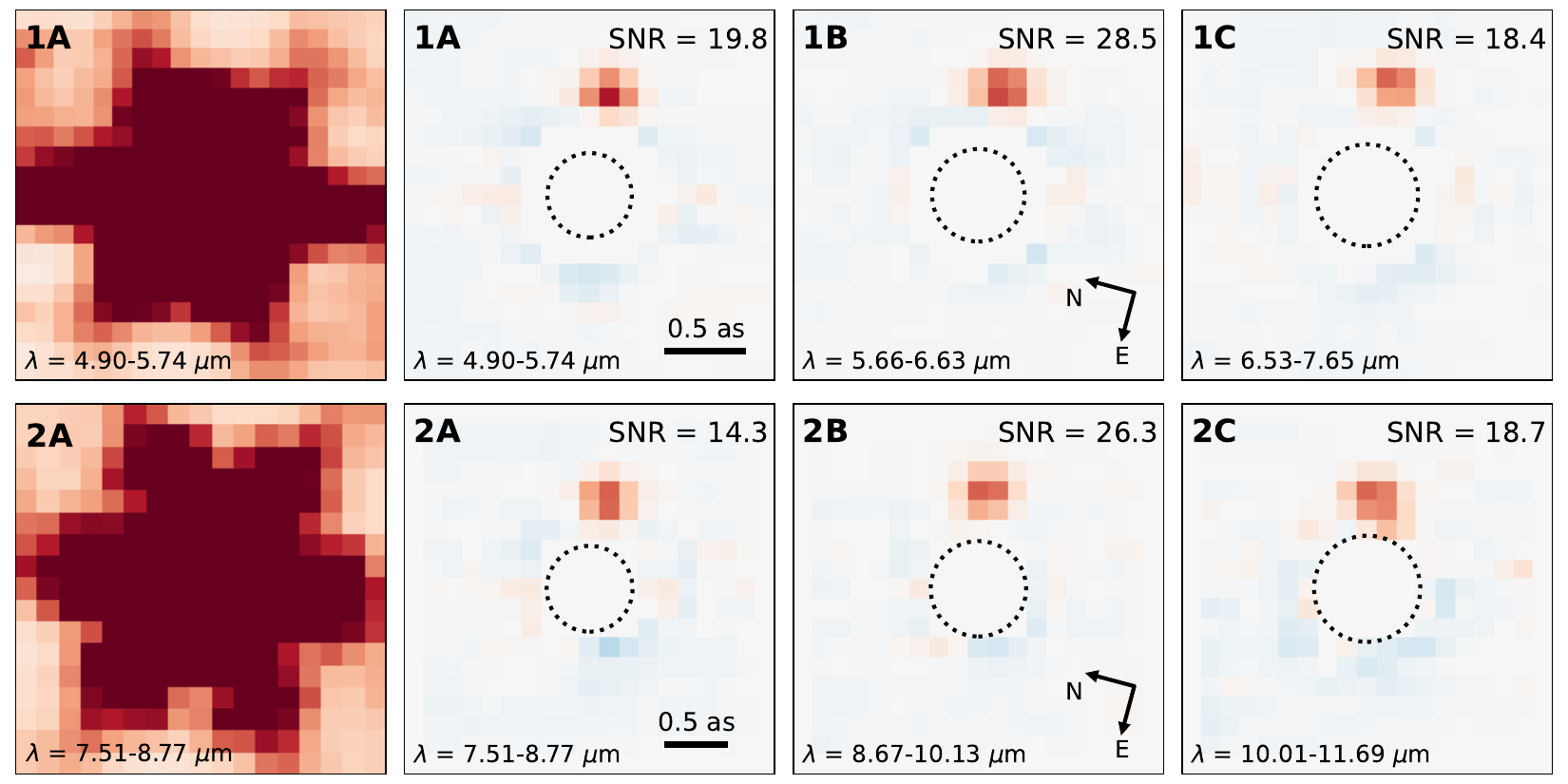}
    \caption{Wavelength-combined images before (first column, only channels 1A and 2A shown) and after (second to fourth columns, channels between 1A and 2C) PSF subtraction. The band is reported on the top left corner of each panel. The corresponding wavelength range is shown on the bottom left corner of each panel, while the SNR of the detection is indicated on the top right corner. The color scale is the same in every image, and the spatial axes follow the MRS IFU internal coordinates. A scale is provided for channels 1 and 2 in the panels of the second column, while arrows pointing North and East in the sky coorindate are reported on panels in the third column. The central dashed circle represents the FWHM of the stellar PSF. With increasing wavelength, the size of the PSF increases and the companion separation falls to $\sim1.5\,\lambda/D$.   }
    \label{fig:residuals_bands}
\end{figure*}

\section{Observations and Data Reduction} \label{sec:obs_red}

\subsection{Observations}\label{sec:observations}
GQ~Lup was observed as part of the general observer (GO) JWST program identifier (PID) 1640 (PI: Banzatti, observation 9) on August 13, 2023. The program was targeting GQ~Lup~A in order to study the inner disk for signs of water emission that would indicate inward pebble drift (see \citealt{Banzatti2023} for more information on PID1640). A point source optimised 4-point dither pattern was used, with 22 groups/integration and 7 integrations (exposure time of 28 mins) for each of the spectral bands of the MIRI Medium Resolution Spectrometer \citep[MRS,][]{Argyriou2023_MRS}. Hence, the observations cover the $4.9-27.9~\mu$m wavelength range, with a spectral resolution ranging from 3700 to 1300. However, in this work we only use channels 1A-2C ($4.9-11.7~\mu$m) as at longer wavelengths the companion can not be spatially resolved. No dedicated background observations were acquired.

The reference star observations used to remove the stellar point spread function (PSF) were taken from a series of commissioning and cycle 1 calibration programs (PIDs: 1050, 1524, 1536, 1538), as well as another star in PID 1640 (RY~Lup, observation 10) that was observed immediately after the GQ~Lup visit. In total 17 reference PSFs were available. These calibration stars have been vetted against binarity or disk emission \citep{Gordon2022}. Additionally they sample a year of JWST capturing potential differences in the state of the wavefront, although this is assumed as a negligible effect at the wavelengths of MIRI. Finally, we selected PSFs with the same dither pattern as our observations (the most common for point sources) in order to sample the PSF in a similar way.

\subsection{Data Reduction}\label{sec:data_reduction}
All data presented in this study were equivalently processed with the \texttt{jwst} pipeline\footnote{pipeline version 1.12.5, CRDS version 11.17, CRDS context jwst\_1141.pmap}. The data were downloaded from the Mikulski Archive for Space Telescopes (MAST) already processed by the \texttt{Detector1Pipeline}, in the form of rate files. Next, the spectroscopic pipeline \texttt{Spec2Pipeline} was applied with the default MRS steps enabled except the residual fringing correction. The spectral cubes were built with the \texttt{Spec3Pipeline} using the drizzle algorithm in IFUALIGN mode, with outlier rejection enabled. The IFUALIGN mode builds the cubes in the MRS internal coordinate system, which is not dependent on the specific V3PA value of each observation, resulting in aligned PSFs \citep{Law2023}. In the end, for each dataset the pipeline provides a single image at each individual wavelength.

We cropped the GQ Lup frames to have the brightest pixel at the center of the frames (new size $19\times19$ and $17\times17$ pixels for channels 1 and 2 respectively, see first column of Fig.~\ref{fig:residuals_bands} for channels 1A and 2A). To avoid interpolation artifacts in the GQ~Lup data, we did not center these images with subpixel precision. 
Conversely, we aligned with subpixel precision all the references to the GQ Lup data using spline interpolation, with the exact position of the star found by fitting a 2D Gaussian function. After masking the central $0\farcs5$, we utilized Principal Component Analysis \citep[PCA][]{AmaraQuanz2012} to extract the dominating features of the PSF. These extracted principal components (PC) form the basis for modelling and removing the GQ~Lup~A PSF, revealing the forming companion. In our analysis we employ 14 PC, but we tested different numbers and found no significant difference in the extracted spectrum for PC$>10$. 

\section{Results}\label{sec:results}
We clearly detected the companion in bands 1A-2C. Figure~\ref{fig:residuals_bands} shows the median collapse along the $\lambda$ axis of the residuals after PSF subtraction in each cube. In addition, the figure reports the signal to noise ratio estimated in the images when using the prescriptions presented in \cite{Mawet2014} with aperture diameters equal to 1~FWHM at the central wavelength of each MRS channel. Until 2C ($\sim11.7~\mu$m) the companion is detected with SNR$\geq$14. Furthermore, up to band 2C, we could recover the signal from GQ~Lup~B in each single frame of the MRS cubes.

Beyond channel 2C ($\lambda>11.7~\mu$m) the companion is located at a separation smaller than 1.5 FWHM of the MRS PSF. Even though the detection at longer wavelengths is possible, it is complicated by a series of factors like significant self-subtraction of the companion flux and higher residual noise that dominates the image. Obtaining a complete spectrum beyond $12~\mu$m is beyond the scope of this paper and is left for future work.

\subsection{Spectral Extraction}
\label{sec:spectral_extraction}
Due to the increasing importance of self-subtraction with increasing wavelengths, we relied on the injection of negative PSFs at the companion location and minimization of the residuals to extract the spectrum of GQ Lup B. The PSF injection is performed at each wavelength before the PCA-PSF subtraction step with an empirical PSF obtained from calibration program PID 1536 (observation 22) at the same wavelength. We chose this dataset as it provides a well-behaved PSF observed at high signal-to-noise ratio (SNR). Moreover, we verified with another dataset (PID 1538, observation 1) that the spectral extraction is independent from the PSF chosen for calibration.
The companion location for the injection is obtained by fitting a 2D Gaussian to the high signal-to-noise median combination of the cube. The residuals are minimized following \cite{Wertz2017} in a circular area of 1.5 FWHM in radius centered on the companion location. Uncertainties are obtained by injecting signals at 180 different position angles and retrieving them with the same algorithm used for GQ~Lup~B. The errorbar resulted from the standard deviation of the differences between the inserted and the retrieved fluxes. In addition to uncertainties, this method allows to correct for biases \cite[e.g.,][]{Stolker2020_miracles, Cugno2024}. For almost every MRS wavelength between 1A and 2C we measure a contrast between 6 and 7 mag with respect to GQ Lup A, with the handful of exceptions being related to bright features like emission lines in the spectrum of the central star. 
The stellar spectrum of the empirical PSF is extracted using the standard \texttt{jwst} pipeline. The multiplication of the stellar spectrum with the contrast at each channel yields accurate absolute photometry, since the spectrum of GQ~Lup~B is correlated to the high signal to noise ($\sim$200) stellar spectrum of the empirical PSF observation.

Finally, some oscillation due to the undersampling of the MRS were still present in the spectrum (see \citealt{Law2023}). To account for this systematic effect for each channel we calculated the root mean square (RMS) of a high-pass filtered spectrum after removing a cubic fit. The filter window (41 wavelength steps) was estimated from the 1C channel, which shows the largest oscillations, by plotting a periodogram and identifying the frequency at which the noise is dominated by random noise. After binning every 100 data points of the spectrum using {\tt spectres} \citep{Carnall2017}, we added in quadrature the RMS just estimated for each channel to the uncertainty. The systematic noise in each MRS channel is a factor 1.6-3.5 (minimum-maximum) the random noise estimated from the data. The low value is consistent with the estimates from \cite{Law2023} in the case of a simple point source extraction, while higher values indicate the presence of more significant systematics likely arising from the high-contrast nature of our observations.

\begin{figure*}[t!]
    \centering
    \includegraphics[width = \textwidth]{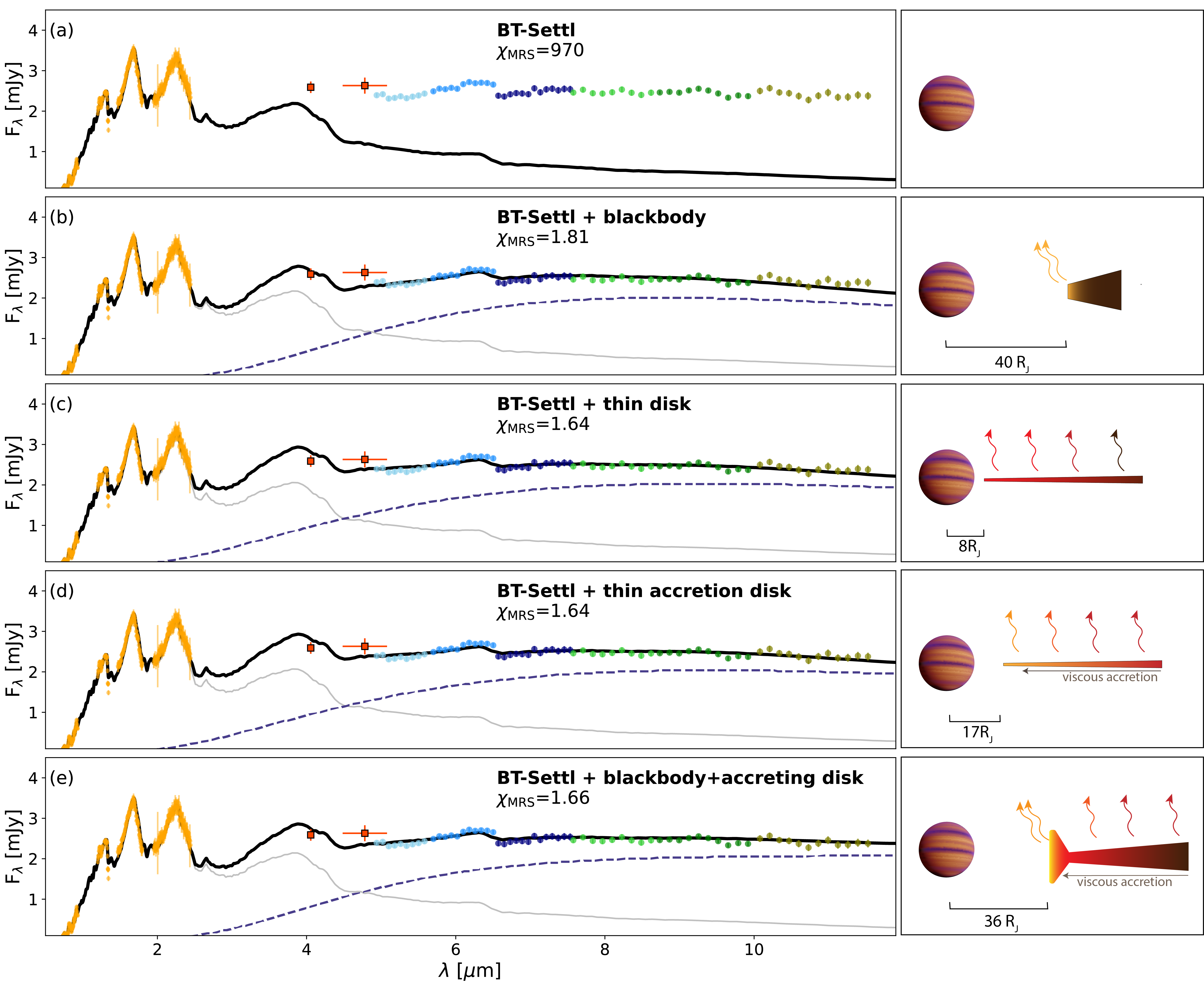}
    \caption{Best fits for the different models considered in this work (left) and schematic representation of the models (right). In the left panels, orange datapoints represent MUSE and SINFONI data, red squares show VLT/NaCo photometries \citep{Seifahrt2007, Stolker2021}, while circles beyond $5~\mu$m show the spectrum extracted from the MRS data (1A-2C). The grey line shows the atmospheric contribution from GQ~Lup~B, the dashed blue line the contribution from the disk and the black thick line shows the overall SED model. Each panel reports the reduced $\chimrs$. {\it Panel a:} best fit model when no disk emission is considered. For the fit, we only used datapoints shortwards of $5~\mu$m, as including the MRS spectra resulted in an unphysical result. Atmospheric emission is not able to account for the strong radiation at $\lambda>5~\mu$m. {\it Panels b-e:} Best fit results for the models that take into account disk emission (model reported on the top right of each panel). In the right panels, the considered scenarios are illustrated. The size of the cavity is reported in each panel, even though the illustration is not to scale. For comparison, the Galilean moons around Jupiter have semi-major axes between 5.9 and $26.3~\RJ$. Brown dwarf artist impression credit: NASA.}
    \label{fig:fits}
\end{figure*}

The obtained spectrum is reported in Fig.~\ref{fig:fits}, together with the results from the fitting procedures from Sections~\ref{sec:SED_fit} and \ref{sec:disk_models}. The atmospheric water feature at $\sim6.6~\mu$m is detected in our spectrum, providing confidence about the robustness of the extraction.


\subsection{SED Fit}
\label{sec:SED_fit}
Since no additional data at $\lambda<5~\mu$m has been published for GQ~Lup~B since \cite{Stolker2021}, we use their same data obtained with VLT/MUSE, VLT/SINFONI and VLT/NaCo. We followed a similar fitting procedure for the atmospheric emission as \cite{Stolker2021}, by interpolating a grid of model spectra and using Bayesian inference for the parameter estimation \citep{Stolker2020_miracles}. In short, we model the atmospheric emission with a BT-Settl model (\citealt{Allard2012}, described by parameters $\Teff$, $\log(g)$, $\Rp$) suffering from extinction $A_V$ (the extinction law from \citealt{Cardelli1989} is considered here). To account for potential flux calibration issues, the $J$ and $H$ band SINFONI data were scaled with free parameters $a_J$ and $a_H$. These are the two spectra that are not anchored to space-based photometric datapoints \citep{Stolker2021}. Initial fits provided poor reduced $\chi^2$, likely due to optimistic errorbars in the MUSE and SINFONI spectra and a model not able to reproduce every feature of the atmospheric emission of the companion. We therefore decided to include error inflation parameters $b_\mathrm{MUSE}$ and $b_\mathrm{SINFONI}$ in the fit, following \cite{Line2015}, so that $\sigma^2=\sigma_\mathrm{INS}^2+10^{b_\mathrm{INS}}$, where $\sigma_\mathrm{INS}$ is the nominal errorbar on the spectra, $b_\mathrm{INS}$ the inflation parameter for MUSE and SINFONI and $\sigma$ is the final errorbar used in the fit. 

We fit the data using {\tt pymultinest} \citep{Feroz2009,Buchner2016}, which allows Bayesian parameter estimation and model comparison. We used 1000 MultiNest live points to explore the parameter space. The list of parameters, together with their prior ranges, can be found in the first and second columns of Table~\ref{tab:values}. To assess the goodness of fit, we calculated the reduced $\rchi^2$ values for the fits 
using the MRS spectrum only (denoted as $\chimrs$). This allows to focus on the MIR wavelengths and evaluate the different models for the circumplanetary disk emission.  

The best fit atmospheric model obtained including all NIR and MIR observations was very poor, suggesting that a more complicated model is required. We then removed the MRS spectrum from the fit, obtaining a good solution at $\lambda<3~\mu$m that is however highly inconsistent with the spectrum extracted for GQ~Lup~B at $\lambda>5~\mu$m (top panel of Fig.~\ref{fig:fits}, see \citealt{Stolker2021}). The third column of Table~\ref{tab:values} reports the median value and the 64\% range (in case of asymmetric posteriors, we report the larger value as the errorbar). The fact that the atmospheric model is unable to match the data beyond $5~\mu$m confirms the presence of IR excess as inferred with NaCo data from \cite{Stolker2021}. The presence of emission lines associated with accretion \citep{Seifahrt2007, Zhou2014, Stolker2021, Demars2023}, the red colors of GQ~Lup~B \citep{Stolker2021} and the presence of this IR excess all point to the presence of a circumplanetary disk surrounding the forming companion.

\begin{table*}[t!]
\centering
\caption{Summary of the parameters used for the SED fit of GQ~Lup~B.}
\def\arraystretch{1.25}
\begin{tabular}{lllllll}\hline
Parameter       & Range         & No disk          & Blackbody      & Geom. thin        & Geom. thin opt.   & Blackbody+    \\
                &               &                  &                & opt. thick disk   & thick accr. disk  & accr. disk    \\\hline
$\Teff$         & $2500-2900$   & $2700\pm18$      & $2718\pm16$    & $2717\pm14$       & $2717\pm14$       & $2719\pm14$  \\ 
$\log(g)$       & $3.3-4.9$     & $4.85\pm0.13$    & $4.86\pm0.12$  & $4.93\pm0.09$     & $4.93\pm0.09$     & $4.91\pm0.11$ \\
$\Rp$ [$\RJ$]   &$0.7-4.0$      & $3.71\pm0.02$    & $3.68\pm0.02$  & $3.60\pm0.03$     & $3.60\pm0.03$     & $3.67\pm0.02$ \\
$A_V$ [mag]     & $0.0-5.0$     & $2.60\pm0.06$    & $2.55\pm0.06$  & $2.43\pm0.08$     & $2.42\pm0.08$     & $2.53\pm0.06$ \\
a$_J$           & $0.5-1.5$     & $1.27\pm0.01$    & $1.26\pm0.01$  & $1.24\pm0.01$     & $1.24\pm0.01$     & $1.26\pm0.01$       \\
a$_H$           & $0.5-1.5$     & $1.07\pm0.01$    & $1.06\pm0.01$  & $1.04\pm0.01$     & $1.04\pm0.01$     & $1.06\pm0.01$      \\
b$_\mathrm{MUSE}$& (*)          & $-31.71\pm0.06$  & $-31.70\pm0.06$& $-31.69\pm0.06$   & $-31.69\pm0.06$   & $-31.70\pm0.06$ \\
b$_\mathrm{SINFONI}$& (*)       & $-31.88\pm0.05$  & $-31.90\pm0.05$& $-31.93\pm0.05$   & $-31.93\pm0.04$   & $-31.91\pm0.04$ \\
$T_\mathrm{BB}$ [K]& $200-900$  & $-$              & $580.9\pm4.2$  & $-$               & $-$               & $616.6\pm11.4$    \\
$R_\mathrm{BB}~[\RJ$]& $1-100$  & $-$              & $27.6\pm0.3$   & $-$               & $-$               & $24.4\pm0.9$\\
$\Rcav~[\RJ$]   & $3-100$       & $-$              & $40.3\pm0.7^\dagger$& $8.2\pm0.8$  & $16.7\pm1.6$      & $35.7\pm1.4^\dagger$\\
$\Rout~[\RJ$]$^\ddagger$& $20-21000$ & $-$         & $-$            & $38.7\pm5.2$      & $79.8\pm10.8$     & $10209\pm4649$\\
$i~[^\circ$]    & $0-90$        & $-$              & $-$            & $71.2\pm2.0$      & $85.6\pm0.5$      & $24.7\pm17.8$ \\\hline
$\chimrs$       &               & 970              & 1.81           & 1.64              & 1.64              & 1.66  \\\hline
\end{tabular}\vspace{0.2cm}
\tablenotetext{*}{Following \cite{Line2015}, we set a prior range defined by $0.01 \times \min(\sigma_\mathrm{INS}^2) \leq 10^b \leq 100 \times \max(\sigma_\mathrm{INS}^2)$, where INS can be either MUSE or SINFONI.}
\tablenotetext{\dagger}{parameter derived according to Eq.~\ref{eq:cavity} and not directly obtained from the fit.}
\tablenotetext{\ddagger}{ The upper bound is defined by the theoretical disk truncation radius, see Sect.~\ref{sec:size}.}

\label{tab:values}
\end{table*}

\subsection{Modelling the disk contribution}
\label{sec:disk_models}

\subsubsection{Blackbody}
We initially attempt at describing the excess emission as being traced by a single blackbody model. In this scenario, the emission comes from the warm dust of the inner disk rim that is well irradiated by the central object. We repeated the atmospheric fit, including the blackbody emission to better fit the MRS spectra. The best-fit is shown in panel (b) of Fig.~\ref{fig:fits}, and its parameters are reported in Table~\ref{tab:values} together with the reduced $\chimrs$. The disk emission is traced by a temperature of $T_\mathrm{BB}=581\pm4$~K.
Using Stefan-Boltzmann's law, and assuming that (i) at 117~au the radiation from GQ~Lup~A is negligible \citep{Stolker2021} and (ii) dust grains are not reflecting any radiation coming from B, we can estimate the radius of the cavity $R_\mathrm{cav}^\mathrm{BB}$ in the circumsecondary disk via 
\begin{equation}
\label{eq:cavity}
    R_\mathrm{cav}^\mathrm{BB} = \sqrt{\frac{L_B}{16~\pi~\sigma~ T_\mathrm{BB}^4}}
\end{equation}
where $L_B$ is the luminosity of GQ~Lup~B and $\sigma$  the Stefan-Boltzmann constant. We obtain that the cavity in GQ~Lup~B has a radius of $R_\mathrm{cav}^\mathrm{BB}\sim40.3\pm0.7~\RJ$ or $\sim0.021$~au. We note that if some of the emission from B is scattered from the disk material, the $L_B$ absorbed by the disk material decreases and the cavity size shrinks.

The $\chimrs$ for the blackbody model provides a significant improvement with respect to the model without the disk, with $\chimrs=1.81$. Despite describing the data reasonably well, this model is simplistic and does not include a radial temperature profile for the disk.

\subsubsection{Geometrically thin optically thick disk}\label{sec:full_disk}
Next, we assumed a geometrically thin optically thick disk model, which has been shown to model the SED of TTauri star at long wavelengths and reproduce their infrared excess \citep[e.g.,][]{Calvet1991, Meyer1997}. We used the temperature prescription from \cite{AdamsShu1986} 
\begin{equation}
T_d (r)= T_* \times \left( \frac{2}{3\pi}\right)^{1/4} \left( \frac{r}{\Rp}\right)^{-3/4}
\end{equation}
and we consider the disk consisting of rings emitting like blackbodies between the inner radius $\Rcav$ and the outer radius $\Rout$, whose emission comes from the reprocessing of the radiation absorbed from the central object GQ~Lup~B. We further consider the disk to be inclined by an angle $i$, so that its total emission is described by
$$F_\mathrm{disk}(\lambda) = \int_{\Rcav}^{\Rout} \frac{2\pi rdr}{D^2}B_\lambda(T_d(r), \lambda)\cos(i).$$

Using this model, the results of the fit suggest that the disk cavity $\Rcav$ is $8.2\pm0.8~\RJ$. Furthermore, the posterior of the disk inclination suggests an inclined ($i\approx71.2\pm2.0^\circ$), and compact ($\Rout\approx38.7\pm5.2~\RJ$) disk. We note, however, that under the geometrically thin assumption a larger disk does not contribute substantially at wavelengths $\lambda = 5-10~\mu$m. As a consequence, this value should be treated with caution, even though a much larger disk can be excluded. 
For this disk model $\chimrs=1.64$, suggesting that contributions from multiple radial separations in the disk seem to slightly improve the fit. 

\subsubsection{Geometrically thin optically thick accreting disk}
Given that accretion tracers like H$\alpha$ and Pa$\beta$ emission lines have been detected in GQ~Lup~B \citep[e.g.,][]{Seifahrt2007, Zhou2014, Demars2023}, we expect additional heating due to viscous accretion. Hence, we added an additional component to the thermal profile of the disk, which is now described by 
\begin{equation}
\label{eq:temp_disk_accr}
T_d (r)= T_* \times \left( \frac{2}{3\pi}\right)^{1/4} \left( \frac{r}{\Rp}\right)^{-3/4} + \left(\frac{G \Mp \dot{M}}{8\pi\sigma r^3} \right)^{1/4}
\end{equation}

Assuming the mass of GQ~Lup~B to be $\Mp=30~\MJ$ \citep{Stolker2021} and its mass accretion rate to be $\dot{M}=10^{-6.5}~\MJ$\,yr$^{-1}$ \citep[and consistent with the lower limits presented in \citealt{Demars2023}]{Stolker2020_pds70, Zhou2014}, we obtain that $L_B>G \Mp \dot{M}/r$ at every separation $r$. Hence, we expect the radiation from B to be the dominant heating source of the disk, rather than viscous accretion. 

Considering accretion increases the disk temperature at every separation. As a consequence, we find a larger cavity compared to the previous model ($\Rcav=16.7\pm1.6~\RJ$). Furthermore, the disk is larger ($\Rout\approx79.8\pm10.8~\RJ$) and more inclined ($i\approx85.6\pm0.5^\circ$), almost in an edge-on configuration. This is inconsistent with the non-detection of polarized light by \cite{vanHolstein2021}. We note that these results strongly depend on two parameters that are only loosely constrained, $\Mp$ and $\dot{M}$. $\chimrs$ is very similar to the one found for the passive disk (see Table~\ref{tab:values}).

\subsubsection{Puffed-up inner wall with an accreting disk}
\label{sec:puffed_rim}
\cite{Dullemond2001} and \cite{Natta2001} suggested that the inner rim of the disk can be `puffed up', due to the strong radiation field coming from the central source. The increased scale height would shadow the disk at larger separations, reducing the disk temperature behind the wall. This hypothesis may partially explain the lack of a silicate feature, as \cite{Calvet1992} suggested that such a feature arises when the disk surface is heated well above the temperature of the disk midplane by the central object's radiation, and why a single blackbody already describes the observed data relatively well. Indeed, if the hot disk rim shadows the outer and colder regions, its emission will dominate the IR-spectrum. For this scenario, we describe the inner disk wall with a single blackbody temperature, and we include a disk at larger separations whose heating is dominated by accretion
\begin{equation}
    \begin{gathered}
        F_{disk}(\lambda) = B_\lambda(T_\mathrm{disk}^\mathrm{BB}(r), \lambda)\times \frac{R_{\mathrm{BB}}^2}{D^2}\, + \\ \quad\quad\quad\quad \int_{\Rcav^\mathrm{BB}}^{\Rout} \frac{2\pi rdr}{D^2}B_\lambda(T_d(r), \lambda)\cos(i)
    \end{gathered}
\end{equation}

where $T_d(r)$ only considers the second term from Eq.~\ref{eq:temp_disk_accr} and $\Rcav^\mathrm{BB}$ is obtained from Eq.~\ref{eq:cavity}. 

This model indicates the cavity is $\Rcav=35.7\pm1.4~\RJ$ and the disk size is unconstrained as the thermal contribution from the shadowed disk is too little at $\lambda<11~\mu$m to provide useful information. Millimeter emission is usually considered to be optically thin \citep[e.g.][]{Ansdell2016, Ansdell2018}, but recent work has found that disks could be optically thick at radio wavelengths \citep{Tripathi2017, Macias2021}, especially for substellar objects \citep[e.g.,][]{Ballering2019, Rab2019}, even in the Lupus star forming region \citep{Xin2023}. The deep ALMA limits presented in \citet[PID 2013.1.00374.S, 0.15 mJy at $870~\mu$m]{MacGregor2017} and \citet[PID 2015.1.00773.S, 0.3 mJy at 1.3~mm]{Wu2017} could provide useful constraints in this regard. Assuming the emission is optically thick, the ALMA non-detection constrain the disk radius to be $<135.3~\RJ$ (95\% quantile). 
Radiative transfer modeling of the disk surrounding GQ~Lup~B should consider the MIR spectrum and the ALMA limits to further constrain the physical properties of the disk. This is left for future work. 


This model provides a better fit than the single temperature blackbody with $\chimrs=1.66$ (see Table~\ref{tab:values}). Even though the improvement in $\chimrs$ is small and should be treated with caution, especially considering that this model has a larger number of parameters. Indeed, the single blackbody scenario (panel (b) of Fig.~\ref{fig:fits}) shows that while the model reproduces very well the emission at $\lambda\approx5-10~\mu$m, at longer wavelengths the emission seems to deviate, needing an additional contribution at colder temperatures. This contribution is provided by the accreting disk beyond $\Rcav^\mathrm{BB}$.


\begin{figure*}[t!]
    \centering
    \includegraphics[width = \textwidth]{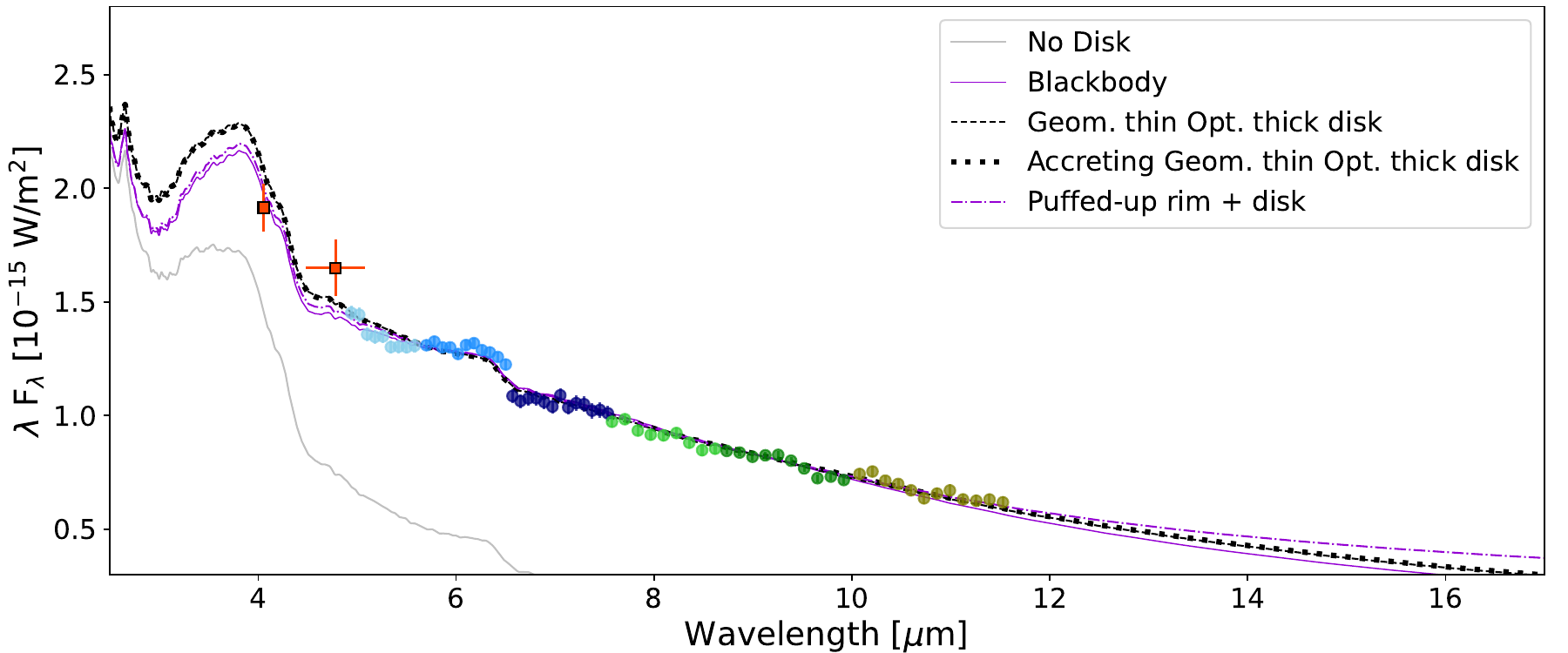}
    \caption{Comparison of the best fit models of the SED of GQ~Lup~B. Red squares show VLT/NaCo photometries \citep{Stolker2021} while circles beyond $5~\mu$m show the spectrum extracted from the MRS data (1A-2C). In black and violet the best fit models for flat and vertically extended disks respectively are shown with different linestyles. $\lambda=2-4~\mu$m and $\lambda>12~\mu$m observations can break the degeneracy and indicate the most suited set of models to describe the observations.}
    \label{fig:cavity}
\end{figure*}

\section{Discussion}
\label{sec:discussion}
\subsection{Circumsecondary disk structure}\label{sec:size}

The SED of GQ Lup B does not reveal silicate features around $9-11~\mu$m, despite a clear feature present in the Spitzer spectrum of the primary \citep{Kessler-Silacci2006}. This likely means that dust grains have already grown to sizes larger than $a_\mathrm{min}\gtrsim5~\mu$m \citep{Woitke2016, Tabone2023}, as for large grain sizes the silicate feature broadens and flattens. \cite{Rilinger2021} found evidence for grain growth in disks surrounding brown dwarfs already at very early stages, including in the Lupus star forming region. Additional factors like small scale height \citep{Szucs2010}, small dust size power index or a large volume fraction of amorphous carbon \citep{Woitke2016} could have led to a fainter or non-existent silicate feature. 

Larger grains settle more efficiently towards the midplane \citep{Dullemond&Dominik2004}, which in turn could increase the optical depth beyond the optically thick limit and lower continuum fluxes, explaining the non-detection of the disk with ALMA. A similar effect could be caused by a low disk flaring. The main difference between these scenarios is the gas temperature: when dust settles, the gas remains warm due to the exposure to the radiation of the central object and emits strong emission lines, while for a flat (low flaring) disk the gas is colder and line emission much fainter \citep{Woitke2016}. The analysis of the spectrum of GQ~Lup~B at higher resolution could provide the necessary information to distinguish between these two scenarios and is left for future work. 

The ratio of the disk luminosity to the stellar luminosity, also known as the fractional disk luminosity, can be used to evaluate the evolutionary status of the disk. Primordial disks are expected to have $L_D/L_*\sim0.1$ as a large fraction of the radiation from the central object is reprocessed and re-emitted by the disk. More evolved disks have much lower fractional luminosities, and debris disks usually present values $\lesssim 10^{-3}$ \citep[e.g.,][]{Bryden2006, Cieza2010}.
After correcting for inclination, our best-fit models provide values for $L_D/L_B$ between 7.7\% (for geometrically thin disks) and 12.3\% (for the puffed-up disk rim), indicating that the disk around GQ~Lup~B is likely at an early stage. We note that this approach has never been tested on disks surrounding very low (planetary) mass objects, and caution is necessary when interpreting these values. 

For stellar multiple systems, it is known that the gravitational interaction between the system's components truncates their radii, resulting in smaller disks \citep[e.g][]{Cox2017, Manara2019, Akeson2019}. For circumplanetary disks, numerical simulations suggest that they might get truncated at about $R_T = 0.3-0.4\,r_\mathrm{Hill}$, where $r_\mathrm{Hill}$ is the Hill radius of the companion \citep[e.g.,][]{MartinLubow2011, ShabramBoley2013}. For GQ~Lup~B, the theoretical upper limit of the disk truncation $R_T$ is $R_T\approx10$~au ($\approx21'000~\RJ$) when using $\Mp=30~\MJ$, $M_*=1~M_\odot$ \citep{MacGregor2017} and $a_p=117$~au \citep{Stolker2021}.

While the fit with vertically extended models does not constrain the extent of the disk (unless the mm emission is optically thick), thin disk models suggest that the dusty disk surrounding GQ~Lup~B is much more compact than its theoretical $R_T$ ($\Rout$ between $38.7$ and $79.8~\RJ$). This might suggest very efficient radial drift of the dust particles, as expected around low-mass brown dwarfs and planetary mass companions \citep[e.g.,][]{Pinilla2013, Zhu2018}. The small sizes, combined with the potential strong settling, could pose a challenge to circumplanetary disk detectability at millimeter wavelengths \citep{Rab2019, Wu2020}.




\subsection{A cavity in a circumsecondary disk}\label{sec:cavity}

The geometry of disks surrounding brown dwarfs is not yet well constrained. While several studies suggest large scaleheights \citep[e.g.,][]{Walker2004, Liu2015, AlvesdeOliviera2013}, other works indicate that these disks are flat and with decreasing scale heights for decreasing mass of the central object \citep[e.g.,][]{Szucs2010}. The choice of the underlying disk model and its vertical structure strongly influences the inference of a cavity in the disk surrounding GQ~Lup~B: while models including a disk wall suggest a cavity up to $\sim37-40~\RJ$ wide, thin disk models suggest a more inclined scenario with a much smaller cavity or even with no cavity at all. Indeed, the dust sublimation radius $R_S$, the separation at which the disk temperature reaches 1400~K and dust sublimates, is $6.6~\RJ$ for an object like GQ~Lup~B. At shorter separations, the dust is expected to be in gaseous form and it does not contribute to the MIR emission, meaning that no dust up to $R_S$ is expected. 
Despite being larger than $R_S$, the cavity radius for the geometrically thin optically thick disk is only 2$\sigma$ away from this value, and thus a disk without cavity can not be excluded. Heating contribution from viscous accretion is expected given the ongoing accretion onto the companion \citep{Zhou2014, Stolker2021, Demars2023}, but its actual impact strongly depends from loosely constrained parameters like companion mass and mass accretion rate. Hence, with the current data it is not possible to definitely prove the circumplanetary disk of GQ~Lup~B has a cavity, and further work is required to discern between the two scenarios.

In the case of wall-dominated disk emission, mechanisms other than dust sublimation need to be at play to create the large gap. 
\cite{Demars2023} found that the Pa$\beta$ line profile of GQ~Lup~B was consistent with magnetospheric accretion (see \citealt{Hartmann2016} for a review)\footnote{We note that the unresolved MUSE H$\alpha$ line showed no evidence of magnetospheric accretion \citep{Stolker2021}, but the line profile might have suffered from oversubtraction from the stellar residuals.}. The cavity for the geometrically thin accreting disk  would be consistent with predictions for this mechanisms, in which the disk inner truncation radius is expected to be a few times $\Rp$ \citep{Hartmann2016}. 
For vertically extended models (blackbody and puffed-up disk) the cavity size is larger than the maximum truncation radius expected from magnetospheric accretion. Even though at this point there is no proof that the cavity is related to the formation of satellites, we note that the orbital semi-major axes of the Galilean moons around Jupiter lays in the range $5.6-26~\RJ$. Hence, they would all fit in the cavity suggested by the vertically extended scenarios. Future radial velocity monitoring of GQ~Lup~B with high spectral resolution instruments like CRIRES and KPIC could investigate the presence of satellites \citep[Horstman et al., in prep.]{Ruffio2023_moons}. 

Figure~\ref{fig:cavity} shows that the degeneracy between disk models can be broken detecting GQ~Lup~B at $L'$ ($\lambda\sim3.8~\mu$m) or at even longer wavelengths. The former can be achieved both from space (e.g., with JWST/NIRCam or JWST/NIRSpec) and from the ground (e.g., with VLT/ERIS), and it would in particular allow to discern if flat or vertically extended disk models better describe the GQ~Lup~B system. Furthermore, more advanced techniques \citep[e.g.,][]{Ruffio2023} could enable the extraction of the spectrum at $\lambda>12~\mu$m from this same MRS dataset, providing crucial evidence for the interpretation of the data and the study of circumplanetary disks.

\subsection{Detection limits for MRS}
\label{detlim}

We investigate the deteciton limits of the MRS data of GQ~Lup. Using the same PSF calibration as in Sect.~\ref{sec:spectral_extraction}, we inserted artificial signals in the dataset where the GQ~Lup~B signal has already been removed, subtract the PSF of the primary and estimate the SNR as in Sect.~\ref{sec:results}. We iteratively adjust the brightness of the artificial signal until it reaches a SNR=5. We repeated this operation for separations spaced by one pixels and for six different position angles, reporting the median detection limit at each separation. 

\begin{figure*}[t!]
    \centering
    \includegraphics[width = \textwidth]{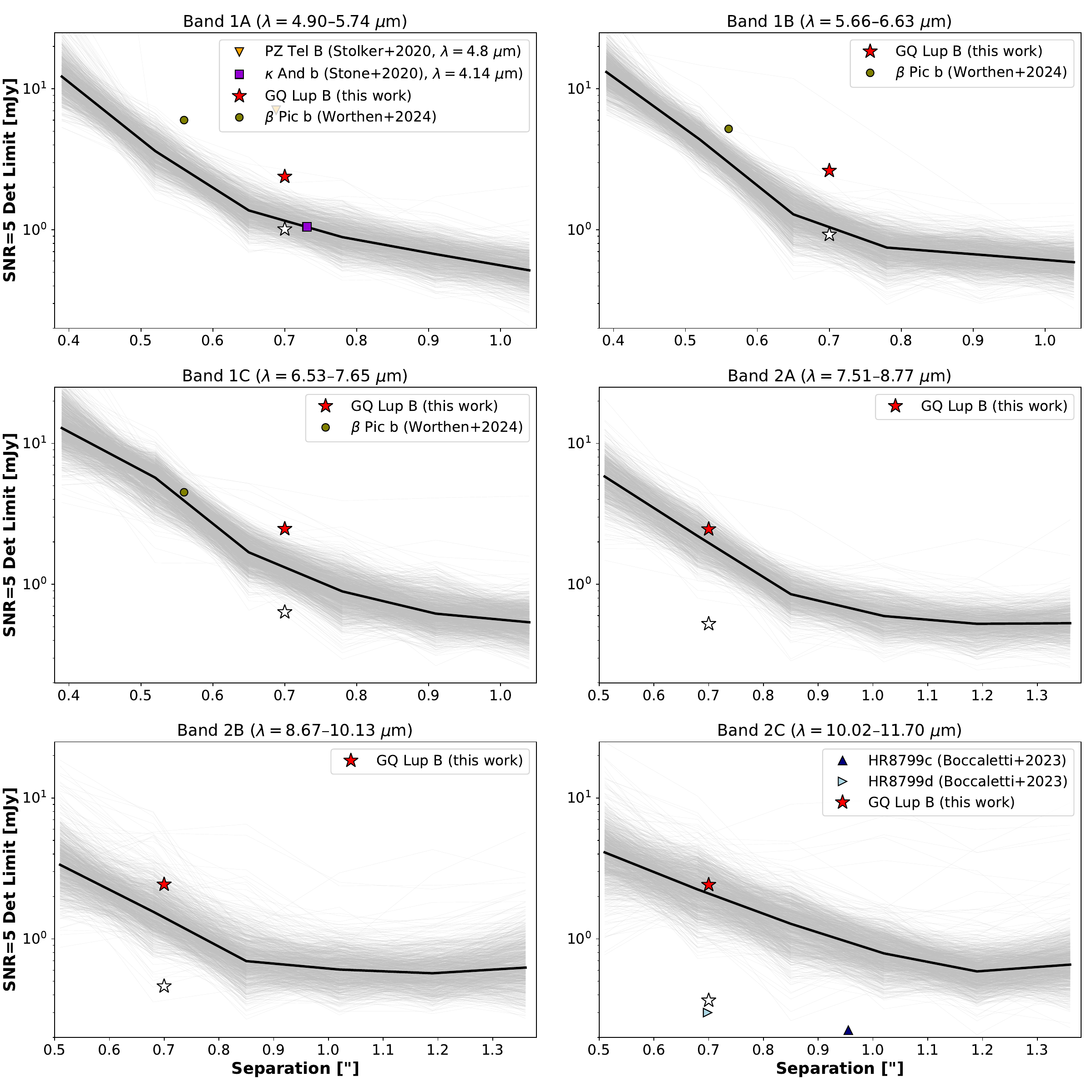}
    \caption{Detection limits of the MRS for bands 1A-2C. Thin grey lines show the limits for each individual MRS wavelength, while black thick lines indicate the median across the channel. For each channel, we report the atmospheric flux of GQ~Lup~B inferred from the NIR spectra as a white star, and the measured flux (atmosphere+CPD) from our MRS observations as a red star. Without the contribution from the circumplanetary disk, GQ~Lup~B would have been detected at only some wavelengths in channels 1A and 1B. In addition, we report MIR fluxes for other known young companions for reference. }
    \label{fig:detlim}
\end{figure*}

Figure~\ref{fig:detlim} reports the detection limits for bands 1A-2C. The limits for each MRS wavelength are plotted in grey and the median at each separation is showed with a thick black line. The star shows the flux of GQ~Lup~B at the center of the MRS band considering only the atmospheric emission (white) and including the CPD contribution (red). For each band, the atmospheric emission of GQ~Lup~B would not be bright enough to be visible in our data (except for some wavelengths in the 1A and 1B channels), but the additional contribution from the CPD makes the companion detectable. 
Additionally, Fig.~\ref{fig:detlim} reports the expected flux of other known companions in several bands \citep[$\beta$~Pic~b, PZ~Tel~B, $\kappa$~And~b, HR8799~cd;][]{Worthen2024, Stolker2020_miracles, Stone2020, Boccaletti2023}. If circumplanetary material exists around other companions, it would manifest as mid-IR excess visible in the MRS bands, favoring their detections. Figure~\ref{fig:detlim} demonstrates the potential provided by the MRS to study young companions and their potential CPD in the mid-IR.

\section{Conclusion}
\label{sec:conclusion}
In this work, we provide the first high-contrast imaging detection of a forming companion with the MIRI/MRS instrument. While \cite{Worthen2024} used simple reference subtraction to remove the stellar PSF of $\beta$~Pic and reveal the b planet and the disk, we relied on a PSF library and PCA to model and subtract the contribution from GQ~Lup~A. We were able to confirm the presence of the mid-IR excess emission from the disk surrounding GQ~Lup~B, and for the first time obtain a $5-12~\mu$m spectrum of a forming low-mass brown dwarf companion and its disk. The data suggest significant grain growth already took place, while additional data will be required to confirm whether the disk hosts a cavity. Future work should employ more elaborate models to determine the physical properties of the disk. 

The approach undertaken in this work can be applied to other circumplanetary disks, potentially around Jupiter-like forming planets. The study of the continuum emission in the MIR allows to study disk structure, accretion and planet formation processes and potentially reveal insights into moon formation around such planetary mass objects.


\begin{acknowledgments}
The authors thank the referee for a useful and constructive feedback that helped improving this manuscript. GC thanks the Swiss National Science Foundation for financial support under grant number P500PT\_206785. PP thanks the Swiss National Science Foundation (SNSF) for financial support under grant number 200020\_200399. This work has been carried out within the framework of the NCCR PlanetS supported by the Swiss National Science Foundation under grants 51NF40\_182901 and 51NF40\_205606. A portion of this research was carried out at the Jet Propulsion Laboratory, California Institute of Technology, under a contract with the National Aeronautics and Space Administration (80NM0018D0004). The authors are grateful for support from NASA through the JWST NIRCam project though contract number NAS5-02105 (M. Rieke, University of Arizona, PI). Some of the data presented in this paper were obtained from the Mikulski Archive for Space Telescopes (MAST) at the Space Telescope Science Institute. The specific observations analyzed can be accessed via \dataset[DOI: 10.17909/3vph-k508]{https://doi.org/10.17909/3vph-k508}. 
\end{acknowledgments}

%

\vspace{5mm}
\facilities{JWST}


\software{{\tt spectres} \citep{Carnall2017}, {\tt pymultinest} \citep{Buchner2016}, {\tt jwst} version 1.12.5 \citep{jwst_pip}
          }

1.tex

\bibliography{gqlup_refs}{}
\bibliographystyle{aasjournal}



\end{document}